\documentstyle[epsf]{article}

 {\par\noindent{\underline{Proof} \quad}}{\hfill$\Box$\bigskip}
 {\par\noindent{\underline{Proof} of the theorem\quad}}{\hfill$\Box$\bigskip}
 {\par\smallskip\noindent{\underline{{\it Remark}} \quad}}{\par\smallskip}
 {\par\smallskip\noindent{\underline{{\it Fact}} \quad}}{\par\smallskip}
 {\par\smallskip\noindent{\underline{{\it Example}} \quad}}{\par\smallskip}
 {\par\smallskip\noindent{{\it Assumotion} \quad}}{\par\smallskip}
 {\par\smallskip\noindent{{\it Condition} \quad}}{\par\smallskip}

\begin{document}
\title{A Clarification on the Detection of Aharanov-Anandan$'$s phase and fault tolerate computation with symmetric SQUID }
\author{Wang Xiang-Bin, Matsumoto Keiji \\
        Imai Quantum Computation and Information project, ERATO, Japan Sci. and Tech. Corp.\\
Daini Hongo White Bldg. 201, 5-28-3, Hongo, Bunkyo, Tokyo 113-0033, Japan}

\maketitle
\begin{abstract} 
We point out that our scheme in manuscript quant-ph/0104127 is misunderstood
by Alexander Blais and Andre-Marie S. Tremblay in their recent e-print quant-ph/0105006.
\end{abstract}
In our scheme(quant-ph/0104127), the state $|\pm>$(eigenstate of $\sigma_y$) will evolve cyclically, with a global
$AA$ phase $\pm \gamma$. In the detection with single qubit, we assume the initial state of $|\psi_0>=|\downarrow>$.
Note this initial state is a linear superposed state of $|\pm>$. 
Explicitly, $|\psi_0>=-\frac{i}{\sqrt 2}(|+>-|->)$. 
Since we have known that in the evolution, state $|\pm>$ will change to $e^{\pm i\gamma}|\pm>$, the initial 
state $|\psi_0>$
will accordingly change into 
$|\psi(2\tau)>=-\frac{i}{\sqrt 2}(e^{i\gamma}|+>-e^{-i\gamma}|->)=-\sin\gamma|\uparrow>-\cos\gamma|\downarrow>$. 
That is to say, $global$ phase $\gamma$ to state $|\pm>$ is $not$ a global phase to state $|\psi_0>$. 
Interference  between $|\psi_0>$ and $|\psi(2\tau)$ $can$ be observed. 
We have never claimed  that we can observe the phase $\gamma$[1] through the 
interference between state $|+>$ and state $e^{i\gamma}|+>$. However, after obtaining the
the pattern by $|<\psi_0|\psi(2\tau)>|^2$, one can deduce the value of $\gamma$, which is the
AA phase shift of state $|+>$. Note $\gamma$ is not the AA phase shift of state $|\psi_0>$. 
This is to say, in order to detect the AA phase $\gamma$ to state $|+>$, we have to observe
the interference pattern by initial state $|\psi_0>$  instead of $|+>$ itself.

To our understanding, the scheme in ref[2] to detect the Berry phase in the single qubit 
case also relies
on the similar interference. There the initial state of qubit is not the eigenstate 
state of the initial Hamiltonian. The qubit will $not$ undergo a closed path in the evolution.  
But the interference pattern is determined by the Berry phase shift to the eigenstate of 
Hamiltonian. Note what is detected there 
is the Berry phase shift to eigenstate of the Hamiltonian rather
than the qubit state itself.  

In the conditional Berry phase part in ref[2], the details are 
not given there. But comparing ref[2] with ref[3], we believe they have first adiabatically
set the qubit to the eigenstate of the Hamiltonian and then operate it. This is different
to the scheme described in the single qubit part in the same article. 
But in general 
it is not necessary to set the initial state of the 
qubit to the eigenstate of the Hamiltonian. Note the final goal is to create the conditional
unitary transformation where no dynamic phase is involved.     

Their note[1] on the fault tolerance part is interesting. 
But we can consider a type of operational error which causes
the random fluctuation on the {\it state evolution path of $|\pm>$}. 
In this case, if the 
area enclosed by the initial state $|\pm>$ is preserved, 
the finally result is not distorted. 
Our scheme is fault tolerate to $certain$  errors in the 
Hamiltonian provided  the area enclosed by the state evolution path of $|\pm>$ is  preserved
under the those errors.  

We believe our scheme for the single qubit case can be realized in laboratory immediately.
We don$'$t know whether our scheme on conditional AA phase shift part
can be realized in laboratory right now, because it is at least as difficult as the realization
of normal C-NOT gate.
 But we
believe that our scheme is 
(theoretically) much closer to the practical use than the old ones.            
\\Wang Xiang-Bin and Keiji Matsumoto\\(Wang X.B. and Matsumoto K.)\\
{\bf Reference}

[1] Alexandre B and Andre-Marie S M, quant/ph-0105006.

[2] C. Falci {\it et al}, Nature {\bf 407}, 355(2000).

[3] Jones {\it et al}, Nature {\bf 403}, 869(2000).
\end{document}